\documentclass{article}
\usepackage[utf8]{inputenc}
\usepackage{geometry}
\newgeometry{hmargin={3cm,3cm}}   
\usepackage{color,amssymb,amsmath,epsfig,ifthen,graphicx,tabularx,xargs}
\usepackage{amsfonts}
\usepackage{amsthm}
\usepackage{algorithm}
\usepackage{algpseudocode}
\usepackage{enumitem,verbatim}
\usepackage{hyperref}
\hypersetup{
    colorlinks,
    linkcolor={red},
    citecolor={blue},
    urlcolor={blue!80!black}
}

\usepackage{cleveref}
\theoremstyle{remark}
\usepackage{mathrsfs}
\usepackage{stmaryrd}
\usepackage{caption}
\usepackage{subcaption}
\usepackage{pdflscape}
\usepackage{authblk}
\usepackage{natbib}
\usepackage{changepage}

\usepackage{newpxmath}

\usepackage[textwidth=2cm,textsize=footnotesize]{todonotes}    

\title{\textbf{Estimand framework development for eGFR slope estimation and comparative analyses across various estimation methods}}
\begin{document}

\author{Tuo Wang\thanks{\url{tuo.wang@lilly.com}} and Yu Du\thanks{\url{du_yu@lilly.com}}}
\affil{Global Statistical Sciences, Eli Lilly and Company, Indiana, United States}

\maketitle
\begin{abstract}
Chronic kidney disease (CKD) is a global health challenge characterized by progressive kidney function decline, often culminating in end-stage kidney disease (ESKD) and increased mortality. To address the limitations such as the extended trial follow-up necessitated by the low incidence of kidney composite endpoint, the eGFR slope—a surrogate endpoint reflecting the trajectory of kidney function decline—has gained prominence for its predictive power and regulatory support. Despite its advantages, the lack of a standardized framework for eGFR slope estimand and estimation complicates consistent interpretation and cross-trial comparisons. Existing methods, including simple linear regression and mixed-effects models, vary in their underlying assumptions, creating a need for a formalized approach to align estimation methods with trial objectives. This manuscript proposes an estimand framework tailored to eGFR slope-based analyses in CKD RCTs, ensuring clarity in defining ``what to estimate" and enhancing the comparability of results. Through simulation studies and real-world data applications, we evaluate the performance of various commonly applied estimation techniques under distinct scenarios. By recommending a clear characterization for eGFR slope estimand and providing considerations for estimation approaches, this work aims to improve the reliability and interpretability of CKD trial results, advancing therapeutic development and clinical decision-making.
\end{abstract}

\section{Introduction} \label{sec:introduction}

Chronic kidney disease (CKD) is a growing global health concern, affecting millions of individuals and imposing a significant burden on healthcare systems worldwiede. CKD is characterized by a progressive decline in kidney function, leading to a variety of adverse health outcomes, including end-stage kidney disease (ESKD), cardiovascular disease, and increased mortality \citep{jha2013chronic, webster2017chronic}. This approach to characterizing kidney function is frequently utilized as the primary objective in randomized clinical trials to demonstrate the efficacy of investigational drugs in improving kidney health. For example, DAPA-CKD trial \citep{heerspink2020dapagliflozin} used a composite endpoint as the primary endpoint. This endpoint was defined as the first occurrence of one of the following events: at least 50\% sustained decline in the estimated Glomerular Filtration Rate (eGFR), the onset of ESKD, or death attributable to renal or cardiovascular causes. However, evaluating therapies for CKD presents a significant challenge in randomized controlled trials (RCTs) using these traditional composite clinical endpoints. Clinical kidney events like ESKD or significant declines in eGFR typically occur in the later states of CKD, often necessitating prolonged follow-up periods to dectect statistically significant differences between treatment and control groups. Such designs risk missing the potential benefits of early intervention, as some therapies may prove more effective when administered in the earlier stages of CKD.

To address this limitation, recent research has increased focus on the eGFR slope as an alternative surrogate endpoint for CKD progression \citep{inker2019gfr,inker2023meta,grams2019evaluating,greene2019performance}. The eGFR slope, defined as the rate of change in eGFR over time, offers a timely and insightful measure that is predictive of long-term clinical outcomes. By capturing the trajectory of kidney function decline, the eGFR slope enables the evaluation of therapeutic interventions at earlier stages of the disease, potentially providing a clearer picture of a drug's efficacy. This shift towards eGFR slope endpoints has garnered growing support from regulatory agencies \citep{levey2020change}, such as the European Medicines Agency (EMA) \citep{ema_gfr_slope} or US Food and Drug Administration (FDA) \citep{thompson2020change}, further solidifying its role in clinical trials.

Despite its growing popularity as a surrogate endpoint, the methodology for estimating the eGFR slope remains a subject of active investigation and debate. A variety of statistical methods have been proposed for eGFR slope estimation, ranging from simple linear regression models to more sophisticated techniques such as linear mixed-effects models, two-slope mixed-effects models, and mixed models for repeated measures (MMRM) \citep{khan2022potential, vonesh2019mixed}. These methods differ significantly in their assumptions about the underlying data, including linearity of the eGFR decline, handling of heteroscedasticity, incorporation of baseline eGFR values, and treatment of varying follow-up durations. As a result, comparisons across studies using different methods can be inconsistent, and the lack of a formalized estimand framework further exacerbates these issues. An estimand framework, as defined by the International Council for Harmonisation of Technical Requirements for Pharmaceuticals for Human Use (ICH), is a conceptual tool that clarifies the target of estimation from a statistical perspective \citep{ich-e9-r1}. ICH E9(R1) emphasizes the importance of clearly defining the estimand, which refers to the specific population parameter to be estimated, before selecting an estimation method. This clarity is crucial for aligning the trial’s objectives with its statistical analyses. Without a standardized estimand framework for eGFR slope, trial results are difficult to interpret consistently across different studies, impeding cross-trial comparisons and potentially undermining the validity of conclusions drawn from these analyses.

This paper aims to fill this gap by proposing a suggested estimand framework for eGFR slope estimation in CKD studies. The framework clarifies ``what to estimate" guiding that subsequent estimation methods are properly aligned with the specified estimand. By establishing a common estimand, a foundation is provided for comparing various estimation methods, both through simulations and real-world data analyses, seeking to provide a robust foundation for future research and clinical applications in the field of CKD RCTs. The ultimate goal is to enhance the reliability and interpretability of eGFR slope estimates, thereby facilitating more informed decision-making in the management of CKD.

The paper is organized as follows. Section~\ref{sec:estimand} outlines a template for establishing the estimand framework in the context of using eGFR slope as the surrogate endpoint. This is followed by Section~\ref{sec:estimation}, which introduces a variety of common estimation approaches for the defined estimand. Section~\ref{sec:simulations} presents extensive simulation studies comparing these methods under various scenarios. Section~\ref{sec:real data example} extends this comparison to real data applications. Finally, the paper concludes with a discussion in Section~\ref{sec:summary}, summarizing key findings and implications for future research.

\section{Estimand definition for eGFR slope analysis} \label{sec:estimand}

\subsection{Estimand framework}
As pointed out in ICH E9 R1, an estimand in clinical research is a well-defined description that encapsulates the specific treatment effect in response to the clinical question of a trial's objective. It anticipates how the outcomes of the same patients might vary under different therapeutic scenarios at a population level. This element is crucial and must be established before initiating a clinical trial, as it steers the study's design and the methods chosen for estimation towards accurate and reliable assessment of the targeted effect. The development of an estimand involves detailed consideration of various attributes, requiring a blend of clinical insight and a strategic approach to addressing intercurrent events in relation to the primary clinical question. 

Attributes required to construct estimand should be clearly defined in the statistical analysis plan in advance of choosing the appropriate estimation methods. This section delves into each of the attributes and provides examples in the context of using eGFR slope as a surrogate endpoint of clinical kidney events. 

\begin{itemize}
    \item \textbf{Population}
    
    In alignment with the clinical question, the estimand framework necessitates a clear and precise definition of the study population. This could encompass the entire cohort of the trial or a specific subgroup, defined by characteristics identified at baseline prior to randomization. This careful delineation ensures that the population studied is directly relevant to the research question and objectives of the trial. For example in DAPA-CKD trial \citep{heerspink2020dapagliflozin}, the population of interest was defined as those ``adults with or without type 2 diabetes who had an eGFR of 25 to 75 mL/min/1.73 m$^2$ of body-surface area and a urinary albumin-to-creatinine ratio (UACR) of 200 to 5000". KDIGO 2024 guideline \citep{stevens2024kdigo} could serve as a good reference when defining patient populations at various stages of CKD. 

    \item \textbf{Variable}

    The variable refers to the specific data or outcome measure collected from each individual in the study that is essential in answering the clinical question of the interest. The value of eGFR is the appropriate variable when using eGFR slope analysis. The chosen formula to define eGFR should be clearly pre-specified according to the recommendation from KDIGO 2024 guideline \citep{stevens2024kdigo} depending on the clinical situations.

    \item \textbf{Treatment}

    This attribute identifies the treatment condition of interest as well as other alternative treatment conditions in comparison. For example in DAPA-CKD trial \citep{heerspink2020dapagliflozin}, participants were randomly assigned to receive dapagliflozin (10 mg once daily) or matching placebo, which defined the treatment conditions being compared in the study. The allowed background therapies should be made clear with this attribute. 

    \item \textbf{Intercurrent Event (ICE)}

    Intercurrent events (ICE) refer to occurrences post treatment initiation that impact the interpretation or existence of outcome measurements relevant to the clinical question. Addressing these events is crucial in defining the treatment effect to be estimated with clarity. In the context of eGFR slope analysis, examples for ICEs include discontinuation of the assigned treatment, use of other therapies prohibited from the study protocol, the occurrence of terminal events like death or ESKD which likely confounds the value of eGFR measurement collected. Treatment policy strategy is typically employed to deal with those ICEs, as is consistent with an intention-to-treat (ITT) principal. With this strategy, the values of eGFR measured are used in the analysis regardless of the occurrence of ICEs. However, terminal events such as death halt the ongoing collection of eGFR values, and events like ESKD can mislead the interpretation of eGFR measurements. To manage these specific scenarios effectively, alternative strategies need to be considered and integrated into the estimand framework, such as composite variable strategy defined in ICH E9(R1).

    \item \textbf{Population-level summary}
    
    Population-level summary summarizes the variable at a population level the comparison of which would capture the treatment effect aligned with the clinical question of interest.

\end{itemize}

\subsection{Weighted average eGFR slope}
Within the estimand framework, we define the population-level summary as the average eGFR slope over a specified time interval. Let $Y(t)$ represent the eGFR at time $t \in [0, \infty)$ and assume $Y(t)$ is differentiable from $(0, \infty)$ with derivative $y(t)$, which represents the instantaneous rate of change in eGFR at time $t$. The weighted average eGFR slope over the interval $[t_1, t_2]$ ($0\leq t_1 < t_2$) is defined as 

\begin{equation}h(t_1, t_2) = \frac{\int_{t_1}^{t_2} w(t)y(t) \textrm{d}t}{ \int_{t_1}^{t_2} w(t) \textrm{d}t}\end{equation}

When the weight function is time-independent, $w(t) = 1$, then the estimand reduces to the average eGFR slope:

\begin{equation}h(t_1, t_2) = \frac{1}{t_2-t_1}\int_{t_1}^{t_2} y(t) dt = \frac{1}{t_2-t_1} \left[Y(t_2) - Y(t_1) \right].\end{equation}

Let $Y^{(a)}(t)$ denote the potential outcome of eGFR at time $t$ under treatment group $a$, where $a=0$ indicates the control group and $a=1$ indicates the treatment group. The estimand is then defined as
$$\mu(t_1, t_2) = E[h^{(1)}(t_1, t_2)] - E[h^{(0)}(t_1, t_2)].$$
If the times are measured in years, $\mu(t_1, t_2)$ is interpreted as the difference in mean rate change in eGFR annually from $t_1$ to $t_2$ between treatment and control group. For example, the commonly used total slope and chronic slope are defined as 
$$\text{total slope} = \mu(0, \tau) = E[h^{(1)}(0, \tau)] - E[h^{(0)}(0, \tau)],$$
$$\text{chronic slope} = \mu(\tau_0, \tau) = E[h^{(1)}(\tau_0, \tau)] - E[h^{(0)}(\tau_0, \tau)],$$
where $\tau$ and $\tau_0$ are pre-specified time points. These estimands capture different aspects of kidney function decline, with the total slope representing the overall effect over the entire follow-up period and the chronic slope focusing on the long-term effect after an initial acute phase. However, the chronic slope depends on post-randomization values, leading to potential selection bias and exaggerating the treatment effects.

It is not hard to see, if $Y(t)$ is a linear function of time $t$ with a constant slope $y(t) = \rho$, then $h(t_1, t_2) = \rho$ for arbitrary $0 \leq t_1 < t_2$ and arbitrary weight function. In this article, we mainly focus on the total slope and for simplicitly define $\mu(\tau) = \mu(0, \tau)$ and $h(\tau) = h(0, \tau)$.

\section{Estimation approaches for eGFR slope analysis} \label{sec:estimation}

In this section, we provide a thorough investigation on commonly used statistical methods to estimate the eGFR slope for the definition in the earlier section. Denote $A$ as the treatment assignment ($A = 0$ for the control group and $A = 1$ for the treatment group). For simplicity, baseline covariates adjustment are not considered, but can be easily incoporated in all the statistical methods discussed in this section. We review the advantages and limitations of each approach, particularly in terms of their underlying assumptions and applicability in different clinical settings.

\subsection{Linear Regression}

Linear regression is the simplest method for estimating the eGFR slope. The model assumes that the eGFR for the $i$th subject at time point $t_{ij}$ can be expressed as: 
$$Y_i(t_{ij} ) = \beta_0 + \beta_1 t_{ij} + \beta_2 A_i + \beta_3 A_i t_{ij} + \epsilon_{ij},$$
where $\epsilon_{ij} \sim \mathcal{N}(0, \sigma^2)$. If the model is correctly specified, the total slope for treatment and control group is $E[h^{(1)}(\tau)] = \beta_1 + \beta_3$, and $E[h^{ (0) }(\tau)] = \beta_1$, respectively. The treatment effect is $\mu(\tau) = \beta_3$, which can be estimated by least squares estimator. In fact, since the slope is constant under linear assumption, we have $\mu(a, b) = \beta_3$. The linear regression assumes the expected value of eGFR slope is the same across all individuals. This assumption of homogeneity ignores  any between-subject variability, potentially leading to biased estimates if there is significant heterogeneity in eGFR decline across individuals.

\subsection{Linear mixed-effects model}

The linear mixed-effects model is a more flexible approach that accounts for both fixed and random effects, allowing for the inclusion of subject-specific variability. The model for the $i$th subject at time point $t_{ij}$ is given by:
$$Y_i(t_{ij}) = \beta_0 + b_{0i} + (\beta_1 + b_{1i}) t_{ij} + \beta_2 A_i + \beta_3 A_i t_{ij} + \epsilon_{ij},$$
where $\epsilon_{ij} \sim \mathcal{N}(0, \sigma^2)$, and $b_{0i}$ and $b_{1i}$ are random intercepts and slopes following a bivariate normal distribution with mean $\boldsymbol{0}$ and covariance matrix $\boldsymbol{\Psi}$. By taking expectation, $E[h^{(1)}(\tau)] = \beta_1 + \beta_3$, $E[h^{ (0) }(\tau)] = \beta_1$, and $\mu(\tau) = \beta_3$, which can be estimated by the approach of restricted maximum likelihood (REML).

In practice, modeling on the change of biomarker is commonly used in industry to assess treatment effects. However, one should be cautious when modeling the eGFR slope using the change in eGFR, $Z_i(t_{ij}) = Y_i(t_{ij}) - Y_i(0)$, as the response variable while including baseline eGFR ,$Y_i(0)$, as a covariate in the model. For example, consider the model 
$$ Z_i(t_{ij}) = \beta_0 + b_{0i} + \gamma Y_i(0) + (\beta_1 + b_{1i}) t_{ij} + \beta_2 A_i + \beta_3 A_i t_{ij}, $$
where $Z_i(t_{ij})$ represents the change in eGFR from baseline at time $t_{ij}$. While this approach adjusts for baseline eGFR to account for potential initial differences across individuals or treatment groups, it fundamentally alters the target estimand. Suppose the observed eGFR data is measured every 6 months in the 3 year period. This model is estimating the chronic slope from  6 months instead of total slope. By modeling changes in eGFR and incorporating baseline as a covariate, the model estimates a parameter that may be misaligned with the clinical question of interest, particularly when the total eGFR slope is the desired outcome.





\subsection{Two-slope linear mixed-effects model}

Both linear regression and linear mixed-effects model assume the population level eGFR is linear. However, this assumption may not hold in certain clinical contexts, such as with drugs in the class of SGLT2 inhibitor, which have been shown to produce an acute, early decline in eGFR followed by a more gradual chronic decline \citep{heerspink2020dapagliflozin,empa2023empagliflozin, perkovic2019canagliflozin}. To address this concern, \cite{vonesh2019mixed} proposed a two-slope linear mixed-effects model , which incorporates a change-point to reflect the different phases of eGFR decline:
$$Y_i(t_{ij}) = \beta_0 + b_{0i} + (\beta_1 + b_{1i}) t_{ij} + (\beta_2 + b_{2i}) \text{max}(t_{ij}-\tau_0, 0) +  \beta_3 A_i + \beta_4 A_i t_{ij} + \beta_5 A \times \text{max}(t_{ij}-\tau_0, 0) + \epsilon_{ij},$$
where $\epsilon_{ij} \sim \mathcal{N}(0, \sigma^2)$, and ($b_{0i}$, $b_{1i}$, $b_{2i}$) are random effects following a multivariate normal distribution with mean $\boldsymbol{0}$ and covariance matrix $\boldsymbol{\Phi}$. $\tau_0$ is a pre-specified changing-point separating the acute and chronic phases of eGFR decline. Then, we have $$E(Y(t_{ij}) | A=1 ) = \beta_0 + \beta_1t_{ij} + \beta_2 \text{max}(t_{ij}-\tau_0, 0) + \beta_4 t + \beta_5 \text{max}(t_{ij}-\tau_0, 0)$$ and $$E(Y(t_{ij}) | A=0)) = \beta_0 + \beta_1t_{ij} + \beta_2 \text{max}(t_{ij}-\tau_0, 0).$$
Then the acute, chronic, total slope for treatment group are 
$E[h^{(1)}(\tau_0)] = \beta_1 + \beta_4$, 
$E[h^{(1)}(\tau_0,\tau)] = \beta_1 + \beta_4 + \beta_2 + \beta_5$, and
$E[h^{(1)}(\tau)] = \beta_1 + \beta_4 + (\beta_2 + \beta_5)(1-\tau_0/\tau)$.
Similarly, the acute, chronic, total slope for control group are 
$E[h^{(1)}(\tau_0)] = \beta_1$,
$E[h^{(1)}(\tau_0,\tau)] = \beta_1 + \beta_2$, and
$E[h^{(1)}(\tau)] = \beta_1 + \beta_2 (1-\tau_0/\tau)$.
Therefore, the treatment effects in acute, chronic, total slope are 
$$\mu(\tau_0) = \beta_4, \quad \mu(\tau_0, \tau) = \beta_4 + \beta_5, \quad \text{ and } \quad \mu(\tau) = \beta_4 + \beta_5 (1-\tau_0/\tau).$$
The regression coefficients can be estimated by REML, and \cite{vonesh2019mixed} suggests to choose the best $\tau_0$ according to Akaike's information criterion (AIC). 

\subsection{Two-stage approach}

The two-stage approach provides an alternative framework for estimating the eGFR slope. In the first stage, an individual slope is estimated for each subject based on a simple linear regression of eGFR on time : 
$$Y_i(t_j) = \beta_{0i} + \beta_{1i}t + \epsilon_{ij}.$$
Each subject will have an estimated slope $\widehat{\beta}_{1i}$ based on least squares estimator. In the second stage, the treatment effect $\mu(\tau)$ can be consistently estimated by comparing the mean slopes between the treatment and control groups:
$$\frac{\sum_{i=1}^{n} \widehat{\beta}_{1i} A_i}{\sum_{i=1}^{n} A_i} - \frac{\sum_{i=1}^{n} \widehat{\beta}_{1i} (1-A_i) }{\sum_{i=1}^{n} (1-A_i) }.$$
Analysis of covariance (ANCOVA) can be used for baseline covariate adjustment. Although this approach is computationally simple, the two-stage approach is particularly sensitive to the issue of missing data, which is common in longitudinal studies.

\subsection{Mixed model repeated measures (MMRM)}

The MMRM model, on a separate note, model time as discrete variables rather than continuous, which is defined as
$$Y_{i}(t_{ij}) = \beta_{0,j} + \beta_{1,j}t_{ij} + \beta_{2,j} A_i t_{ij} + \epsilon_{ij},$$
where $\epsilon_{i1}, \cdots, \epsilon_{iJ}$ follows a multivariate normal distribution with mean $\boldsymbol{0}$ and covariance matrix $\boldsymbol{\Sigma}$. In MMRM model, the random effects are implicitly represented in the covariance matrix, allowing for the specification of different covariance structures (e.g., unstructured, compound symmetry, autoregressive) to best match the pattern of correlations observed in the data. This flexibility makes MMRM particularly useful for analyzing data with missing observations, as it can more accurately model the underlying correlation structure without the need for imputation. Moreover, MMRM focuses on the marginal mean of the outcome variable, providing direct estimates of the mean response at each time point, adjusted for baseline covariates. Specifically, ${\mu}(t_j, t_k) = {\beta}_{2,k} - {\beta}_{2,j}$ with $\beta_{2,0} = 0$. For example, the total slope from $0$ to $t_J$, $\mu(t_J)$, can be estimated by $\widehat{\mu}(t_J) = \widehat{\beta}_{2,J}$ based on REML. Compared to linear mixed-effects models, MMRM provides a more flexible framework for situations where the linearity assumption for eGFR decline may be violated. By treating time as a series of discrete points, MMRM does not impose a strict linear structure on the relationship between eGFR and time, which can yield more accurate estimates when the trajectory of kidney function is non-linear. However, this flexibility comes with certain trade-offs. When the assumption of linearity is approximately correct, MMRM may be less powerful than linear mixed-effects models, which explicitly model time as a continuous variable.
 
%

\section{Simulations} \label{sec:simulations}

\subsection{Data generation}

In this section, we conducted simulation studies to evaluate the performance of various statistical methods described in Section 3 for estimating the average eGFR slope. We considered four settings for generating the true eGFR values with repeated measurements every 6 months for 3 years. The treatment effect of interest is defined as the difference in the average slope, or total slope, from baseline to $\tau = 3$ years.

\textbf{Setting 1: Linear eGFR decline} 
In this scenario, we assume that the true eGFR is a linear trajectory over time for both treatment and control groups. The true eGFR values are simulated from 
$$Y_i(t_{ij}) = \beta_{0a} + b_{0i} + (\beta_{1a} + b_{1i})t_{ij} + \epsilon_{ij},$$
where $\beta_{0a}$ and $\beta_{1a}$ ($a=c,t$) are the fixed effects for treatment and control group, respectively, $b_{0i}$ and $b_{1i}$ are the random effects following a bivariate normal distribution, and $\epsilon_{ij} \sim \mathcal{N}(0, 1^2)$ is the residual error. Specifically, we let $\beta_{0c} = \beta_{0t} = 47.5$, $\beta_{1c} = -2.25$, $\beta_{1t} = -1.5$, $b_{0i} \sim \mathcal{N}(0, 17.5^2)$, $b_{1i} \sim \mathcal{N}(0, 1^2)$, and $\text{correlation}(b_{0i}, b_{1i}) = 0.5$. The baseline eGFR of 47.5 mL/min/1.73m$^2$ are selected based on some real CKD clinical trials \citep{perkovic2024effects}. Under this set-up, the true treatment effect in total slope is $0.75$ mL/min/1.73m$^2$ per year.

\textbf{Setting 2: Nonlinear eGFR decline} 
In this scenario, the true eGFR is a nonlinear function of time $t$ for both treatment and control group. The true eGFR is simulated from 
$$Y_i(t_{ij}) = \beta_{0a} + b_{0i} + (\beta_{1a} + b_{1i})\text{log}(5t_{ij} + 1) + \epsilon_{ij}.$$
The values of the parameters are identical to setting 1 except we set $\beta_{1c}=-2.31$. Under this set-up, the average slope for treatment and control groups are $\beta_{1t}\text{log}(16)/3$ and $\beta_{1c}\text{log}(16)/3$, respectively, yielding a treatment effect in total slope of approximately $0.75$ mL/min/1.73m$^2$ per year..

\textbf{Setting 3: Negative acute effect} 
This setting assumes there is negative acute effect of eGFR caused by the treatment for the first 6 months (0.5 year). Specifically, the eGFR values for treatment group is simulated from 
$$ Y_i(t_{ij}) = \begin{cases} \beta_{0t} + b_{0i} + (\beta_{1t} + b_{1i})t_{ij} & \text{if} \quad t_{ij} \leq 0.5 \\
 \beta_{0t} + b_{0i} + (\beta_{1t} + b_{1i})t_{ij} + (\beta_{2t} + b_{2i}) \text{max}(t_{ij}-0.5, 0)  & \text{if} \quad t_{ij} > 0.5 \end{cases} ,$$
where $\beta_{0t} = 47.5$, $\beta_{1t} = -4$, $\beta_{2t} = 3$, $b_{0i} \sim \mathcal{N}(0, 17.5^2)$, $b_{1i} \sim \mathcal{N}(0, 2^2)$, and $b_{2i} \sim \mathcal{N}(0, 2^2)$. For simplicity, we assumed the random effects are mutually independent. For control group, since there is no acute effect, the true eGFR values are simulated based on setting 1 with the only difference being $\beta_{1c} = -2.25$. Under this set-up, the average slope for treatment group is $\beta_{1t} + \beta_{2t}(3-0.5)/3$, while the average slope for control group is $\beta_{1c}$. Therefore, the treatment effect in total slope is $0.75$ mL/min/1.73m$^2$ per year..

\textbf{Setting 4: Positive acute effect} 
This setting assumes there is positive acute effect of eGFR caused by the treatment for the first 6 months (0.5 year). The eGFR values for treatment group is simulated based on setting 3 with $\beta_{0t} = 47.5$, $\beta_{1t} = 2.5$, $\beta_{2t} = -4.5$. For control group, the true eGFR values are simulated based on setting 1 with the only difference being $\beta_{1c} = -2.5$. Similar to setting 3, under this set-up, the treatment effect in total slope is $1.25$ mL/min/1.73m$^2$ per year..

The trends of eGFR values under setting 1-4 are summarized in Figure 1 and the true average slopes for treatment and control group under setting 1-4 are summarized in Table 1. We only considered independent censoring with censoring time $C \sim \text{Uniform}[0.5, 3]$. For each replicate, we simulated 400 subjects with 200 treated subjects and 200 control subjects. The treatment effect is then estimated by the statistical methods described in Section 3. We replicated the simulation 2,000 times, and the results, including the estimation, bias, standard deviation (SD) and root mean squared error (RMSE) are summarized in Table 2.

\begin{figure}[ht]
\centering
\includegraphics[scale=0.7]{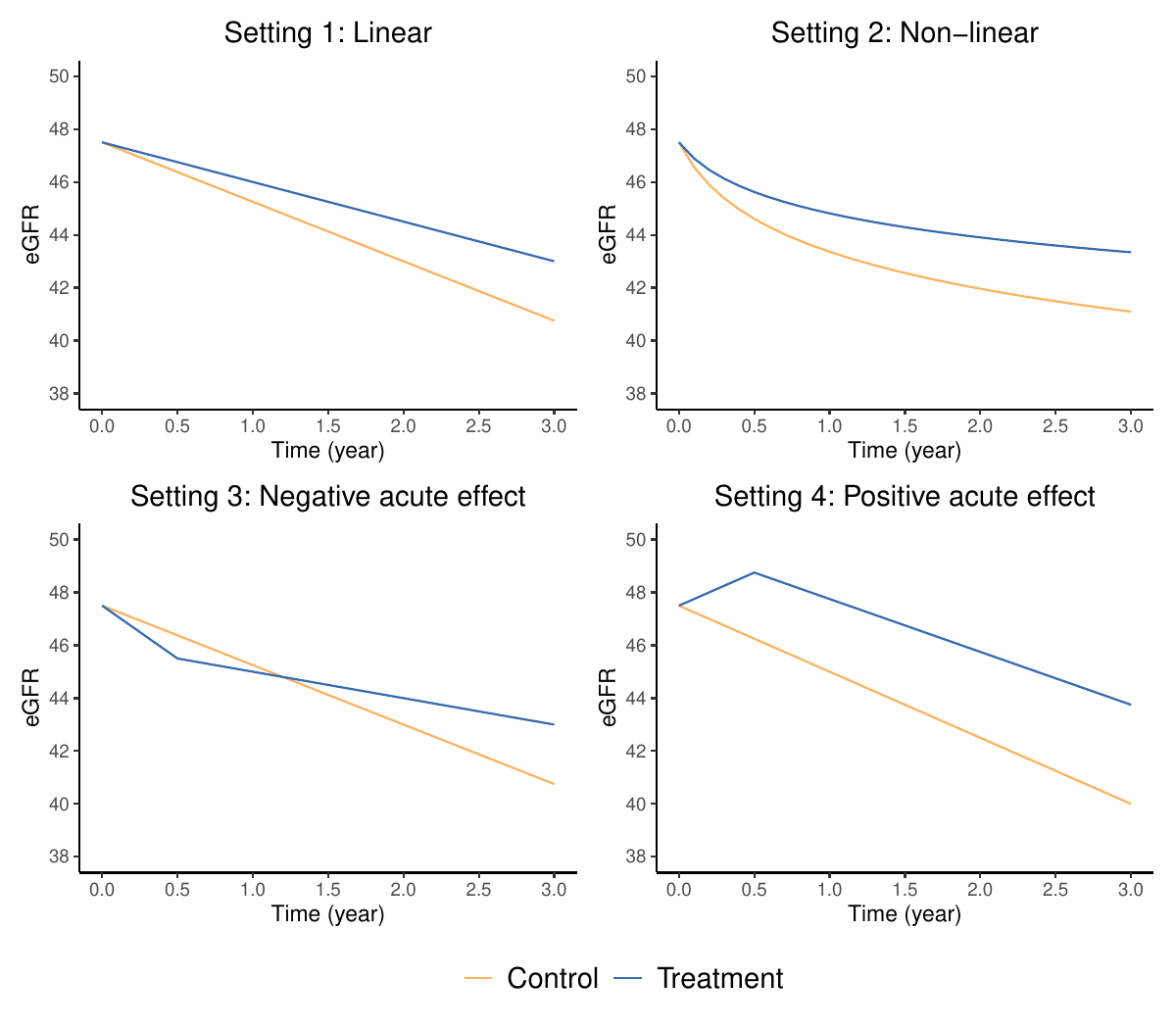}
\caption{ eGFR values as a function of time by treatment group under setting 1-4}
\label{fig:fig1}
\end{figure}

\begin{table}[ht]
\centering
\begin{tabular}{cccc} \hline  \noalign{\medskip}
        & \multicolumn{3}{c}{Average slope from baseline to year 3}  \\  \noalign{\medskip} \cline{2-4}  \noalign{\medskip}
Setting & Treatment          & Control          & Difference         \\ \hline  \noalign{\medskip}
1       & -1.50              & -2.25            & 0.75              \\
2       & -1.39              & -2.14            & 0.75              \\
3       & -1.50              & -2.25            & 0.75              \\
4       & -1.25              & -2.50            & 1.25              \\  \noalign{\medskip} \hline
\end{tabular}
\caption{True average slope under Setting 1-4} 
\label{tab:tab1}
\end{table}

\subsection{Simulation results}

We compared the performance of five estimation approaches described in Section 3. The results are presented in Table 4. 

When the assumption of a linear eGFR decline is true (Setting 1), all five models provide unbiased estimates of the treatment effect. However, the linear mixed-effects model (LME) demonstrates the highest efficiency, yielding the smallest standard deviation (SD) and root mean squared error (RMSE). This is because the LME explicitly models the linear trajectory of eGFR decline while accounting for subject-specific variability via random effects. 

In cases where the eGFR trajectory is nonlinear (Setting 2), the performance of the models diverges. The linear mixed-effects models (LME and Two-slope LME), as well as the linear model (LM), assume a constant slope over time, and therefore produce biased estimates when the true relationship is nonlinear. In contrast, the MMRM approach, which does not assume linearity, remains robust and produces unbiased estimates in this setting. The MMRM allows for the flexibility of estimating the marginal mean at each time point without imposing a rigid structure on the time-eGFR relationship.

Thus, when there is reason to believe that the eGFR decline may not be linear—whether due to the nature of the disease or the mechanism of action of the treatment—MMRM is a robust approach. While the two-slope LME model attempts to capture nonlinearity via a change-point (as in Setting 3), it does not fully accommodate continuous nonlinear trends and may result in biased estimates in settings like Setting 2.

Settings 3 and 4 represent scenarios where there is an acute effect of treatment during the early stage (i.e., the first six months). The results show that when there is a negative or positive acute treatment effect, the linear mixed-effects model (LME) produces biased results. In contrast, both the two-slope LME and MMRM models produce unbiased estimates of the treatment effect in these settings. The two-slope LME model explicitly models the different slopes before and after the acute effect by introducing a change-point, making it more efficient than MMRM in settings with acute effects. The MMRM, while flexible, does not account for abrupt changes in slope in a structured way, but still provides unbiased estimates by estimating the mean response at each time point without assuming a constant slope.

Given these results, when acute effects are anticipated—such as in the case of drugs like SGLT2 inhibitors, which have a well-documented early eGFR dip—using the two-slope LME model would be preferred. It is more efficient than MMRM in these scenarios because it specifically models the change in slope at the acute phase. However, it may be challenging to select the correct change-point in practice. Therefore, MMRM remains a valid alternative but may be less efficient due to its reliance on time-point-specific marginal means rather than explicit slope changes.

In summary, the choice of estimation method should be guided by the underlying assumptions about the eGFR trajectory. In general, the linear model and two-stage model are not recommended to use in practice. In practice, when the data exhibit nonlinearity or acute effects, more flexible models like MMRM or the two-slope LME should be considered over the standard LME or simpler methods like linear regression.

\begin{table}[ht]
\begin{adjustwidth}{-2cm}{-2cm} 
\centering
\resizebox{1.2\textwidth}{!}{
\begin{tabular}{cccccccccccccccc} \hline 
         & \multicolumn{3}{c}{LM} & \multicolumn{3}{c}{LME} & \multicolumn{3}{c}{Two-slope-LME} & \multicolumn{3}{c}{Two-stage} & \multicolumn{3}{c}{MMRM} \\  \cline{2-16} 
Settings & Bias   & SD   & RMSE   & Bias    & SD   & RMSE   & Bias      & SD      & RMSE     & Bias      & SD     & RMSE     & Bias    & SD    & RMSE    \\ \hline  
1 & -0.006 & 0.329 & 0.329 & 0.001 & 0.131 & 0.131 & 0.001 & 0.133 & 0.133 & 0.000 & 0.172 & 0.172 & -0.002 & 0.162 & 0.162 \\
2 & 0.046 & 0.348 & 0.351 & 0.145   & 0.149 & 0.208 & 0.072 & 0.130 & 0.148 & 0.415  & 0.235 & 0.477 & 0.000 & 0.144 & 0.144   \\
3 & 0.030 & 0.742 & 0.742 & -0.339  & 0.222 & 0.405 & -0.006 & 0.232 & 0.232 & -0.635 & 0.239 & 0.679 & -0.005 & 0.260 & 0.260   \\
4 & -0.064 & 0.777 & 0.779 & 0.478 & 0.226 & 0.529 & 0.004 & 0.231 & 0.231 & 0.941 & 0.247 & 0.973 & 0.002 & 0.259 & 0.259 \\  \hline
\multicolumn{16}{l}{Notes: LM: linear model; LME: linear mixed-effects model; Two-slope-LME: Two slpoe lienar mixed-effects model; MMRM: Mixed model } \\
\multicolumn{16}{l}{repeated measure. Bias is the difference between true value and its estimator; SD: standard deviation; RMSE: root mean squared error.}
\end{tabular}}
\caption{Simulations results for different approaches under sample size $n=400$. } 
\end{adjustwidth}

\end{table}


\section{Real Data Example} \label{sec:real data example}

The AWARD-7 study was a randomized, open-label, phase 3 clinical trial designed to evaluate the efficacy and safety of dulaglutide, a glucagon-like peptide-1 (GLP-1) receptor agonist, in patients with type 2 diabetes and moderate-to-severe chronic kidney disease (CKD) \citep{tuttle2018dulaglutide}. The primary objective of the study was to assess the glycemic control achieved by dulaglutide compared to insulin glargine, with both treatments administered alongside prandial insulin lispro. AWARD-7 enrolled patients with an eGFR of 15 to 59 mL/min/1.73 m², reflecting moderate-to-severe CKD. The study demonstrated that dulaglutide not only provided effective glycemic control but also had a beneficial effect on preserving kidney function compared to insulin glargine. Specifically, dulaglutide groups exhibites a smaller decline in eGFR over time compared to the insulin glargine group.

We reanalyzed the AWARD-7 data on the eGFR slope using different estimation approaches described in Secion 3. The estimand of interest was the total slope from baseline to $\tau = 1$ year. Figure 2 presents the least squares (LS) means of eGFR at different points, estimated using the MMRM. For the two-slope linear mixed effect model, we specified week 8 as the change-point to account for potential early acute effects.

Overall, all the models indicates that dulaglutide 1.5mg and 0.75mg slowed the decline of kidney function compared to insulin glargine, though none of the treatment effects are statistically significant. However, the magnitude of the estimated treatment effect varied considerably across the different estimation methods. Notably, the two-slope linear mixed-effects model and the MMRM produced similar estimates of the treatment effect. These results reinforce the importance of aligning appropriate estimation methods to estimand to accurately assess treatment effects on eGFR slope in clinical trials involving CKD patients.

\begin{figure}[ht]
\centering
\includegraphics[scale=0.7]{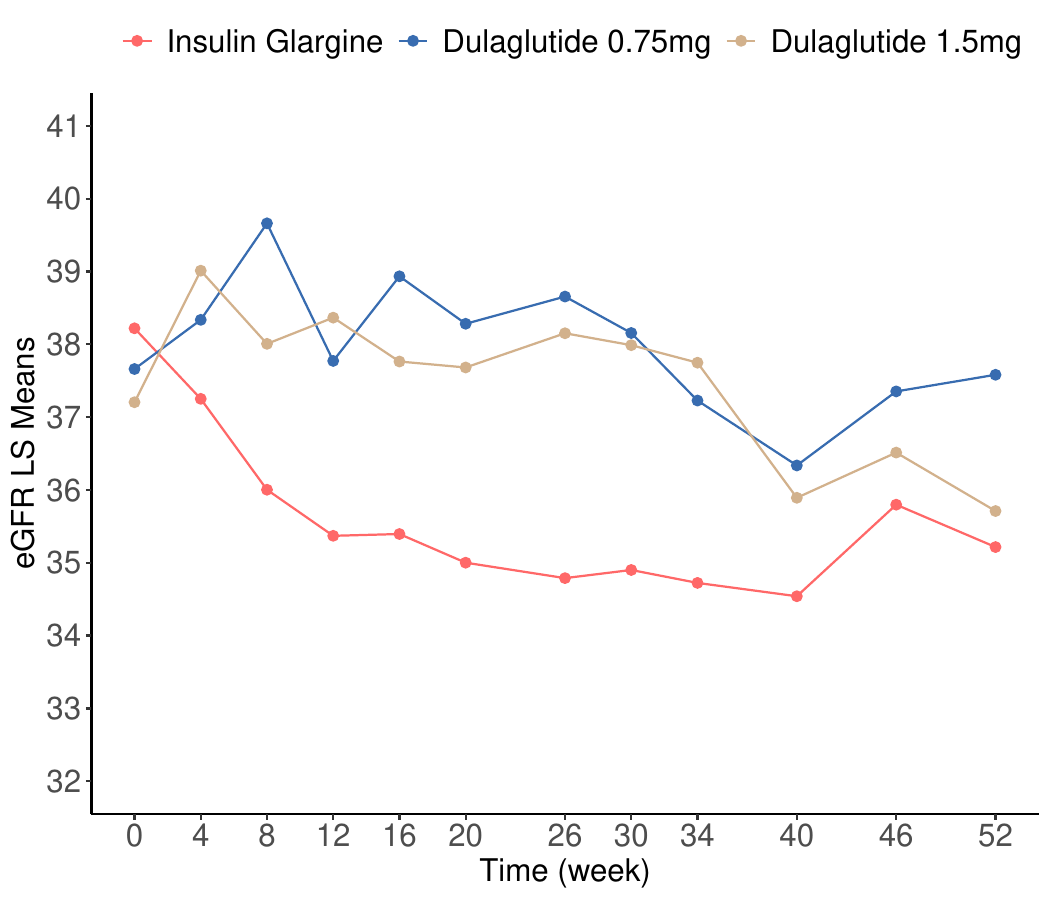}
\caption{ Least squared means of eGFR}
\label{fig:fig2}
\end{figure}

\begin{table}[ht]
\centering
\begin{tabular}{clccc}
\hline
\\
Method               & Week 0 - Week 52 & Dulaglutide 1.5mg & Dulaglutide 0.75mg & Insulin Glargine    \medskip  \\ \hline \\ 
 LM                   & Slope            &      -1.49 (0.89)   &       -1.01 (0.88) & -1.65 (0.87)         \\ 
\multicolumn{1}{l}{} & Difference       &      0.16 (-2.28, 2.59)   &   0.64 (-1.79, 3.06)  & \multicolumn{1}{l}{} \medskip  \\  

LME                  & Slope            &       -1.99 (0.76)   &   -1.41 (0.75)   & -2.14 (0.74)         \\
\multicolumn{1}{l}{} & Difference       &      0.15 (-1.93, 2.23)   &    0.73 (-1.33, 2.80)                & \multicolumn{1}{l}{} \medskip \\

Two-slope-LME        & Slope            &         -1.45(0.79)           &       -0.77(0.78)             &      -3.17 (0.77)           \\
\multicolumn{1}{l}{} & Difference       &         1.71 (-0.45, 3.87)    &     2.40 (0.25, 4.54)               & \multicolumn{1}{l}{} \medskip  \\

Two stage            & Slope            &        -1.83 (1.15) & -1.73(1.16)  & -2.49 (1.13)     \\
\multicolumn{1}{l}{} & Difference       &      0.66 (-2.42, 3.74)  & 0.76 (-2.34, 3.85)    & \multicolumn{1}{l}{} \medskip \\

MMRM                 & Slope            &      -1.49 (1.17)  &     -0.08 (1.16) &  -3.00 (1.14) \\
\multicolumn{1}{l}{} & Difference       &     1.51 (-1.70, 4.73)   &      2.93 (-0.28, 6.13)   & \multicolumn{1}{l}{} \medskip  \\ \hline
\multicolumn{5}{l}{Notes: for slope, the estimates (SD) are summarized; For difference between groups, the }  \\
\multicolumn{5}{l}{ estimates (95\% CI) are summarized. } 
\end{tabular}
\caption{AWARD-7 eGFR slope results under different approaches. } 
\label{tab:tab3}
\end{table}


\section{Summary and Discussion} \label{sec:summary}

In this paper, we introduce an estimand framework tailored specifically for eGFR slope analysis. The framework defines the treatment effect as the difference in average eGFR slopes between treatment and control groups over a specified time period. By clearly defining "what to estimate," this approach guides that the choice of statistical methods is aligned with the clinical objectives of the study. Moreover, it facilitates consistent interpretation of treatment effects across different trials. The application of the estimand framework addresses a key gap in current CKD research, where inconsistencies in slope estimation methods can lead to challenges in cross-study comparisons. By formalizing the estimand as the average slope over a specified time period, our framework ensures that all estimation methods are focused on answering the same clinical question, even when different statistical approaches with different model assumptions are applied. This contribution is particularly valuable in light of increasing regulatory emphasis on well-defined estimands, as reflected in the ICH E9(R1) guidelines. Our framework encourages transparency in analysis and supports a more standardized approach to interpreting eGFR slope as a surrogate marker in CKD trials.

Through simulation studies, we compared the performance of five commonly used estimation approaches. Four distinct scenarios were considered, representing different clinical trajectories: linear eGFR decline, nonlinear decline, and decline with both positive and negative acute treatment effects. The results demonstrated that the linear mixed-effects model is most efficient when the eGFR trajectory is linear, producing unbiased estimates with the smallest RMSE. However, in the presence of nonlinear eGFR decline, the linear mixed-effect models are targeting at the time-dependent weighted average slope, resulting biased estimates to the average slope. The MMRM model, on the other hand, provided unbiased estimates to the average slope. For scenarios involving acute treatment effects, both the two-slope LME and MMRM produced unbiased results, with the two-slope LME showing greater efficiency due to its ability to model different phases of eGFR decline. Simpler methods, such as linear models and two-stage approaches, resulted in biased estimates across most settings and are not recommended for use in practice.

A key limitation of our simulation studies is that we only considered independent censoring. Informative censoring, which often arises in CKD trials due to competing risks such as ESKD or death, could potentially affect the performance of the estimation methods. In the context of informative censoring, the shared parameter method proposed by Vonesh et al. (2019) that simultaneously account for time-to-event outcomes and longitudinal eGFR data may also provide a more comprehensive approach to analyzing CKD progression in clinical trials \citep{vonesh2019mixed,vonesh2022biased}.  Additionally, the development of more advanced nonlinear models, particularly those that can accommodate individual-level variability in both the acute and chronic phases of eGFR decline, may further improve the accuracy and interpretability of treatment effect estimates. 

As the regulatory acceptance of eGFR slope as a surrogate endpoint continues to grow, there is a need for consensus on best practices for its estimation. Establishing a standardized framework for defining estimand and choosing appropriate methods based on trial-specific characteristics (e.g., expected trajectory, patient population, duration) will be critical for ensuring consistency and comparability across studies.

\clearpage

\bibliographystyle{apalike}
\bibliography{ms}

\begin{thebibliography}{}

\bibitem[{European Medicines Agency}, 2023]{ema_gfr_slope}
{European Medicines Agency} (2023).
\newblock Qualification opinion on gfr slope as a validated surrogate endpoint in rcts for ckd.

\bibitem[Grams et~al., 2019]{grams2019evaluating}
Grams, M.~E., Sang, Y., Ballew, S.~H., Matsushita, K., Astor, B.~C., Carrero, J.~J., Chang, A.~R., Inker, L.~A., Kenealy, T., Kovesdy, C.~P., et~al. (2019).
\newblock Evaluating glomerular filtration rate slope as a surrogate end point for eskd in clinical trials: an individual participant meta-analysis of observational data.

\bibitem[Greene et~al., 2019]{greene2019performance}
Greene, T., Ying, J., Vonesh, E.~F., Tighiouart, H., Levey, A.~S., Coresh, J., Herrick, J.~S., Imai, E., Jafar, T.~H., Maes, B.~D., et~al. (2019).
\newblock Performance of gfr slope as a surrogate end point for kidney disease progression in clinical trials: a statistical simulation.

\bibitem[Group, 2023]{empa2023empagliflozin}
Group, E.-K.~C. (2023).
\newblock Empagliflozin in patients with chronic kidney disease.
\newblock {\em New England Journal of Medicine}, 388(2):117--127.

\bibitem[Heerspink et~al., 2020]{heerspink2020dapagliflozin}
Heerspink, H.~J., Stef{\'a}nsson, B.~V., Correa-Rotter, R., Chertow, G.~M., Greene, T., Hou, F.-F., Mann, J.~F., McMurray, J.~J., Lindberg, M., Rossing, P., et~al. (2020).
\newblock Dapagliflozin in patients with chronic kidney disease.
\newblock {\em New England Journal of Medicine}, 383(15):1436--1446.

\bibitem[Inker et~al., 2023]{inker2023meta}
Inker, L.~A., Collier, W., Greene, T., Miao, S., Chaudhari, J., Appel, G.~B., Badve, S.~V., Caravaca-Font{\'a}n, F., Del~Vecchio, L., Floege, J., et~al. (2023).
\newblock A meta-analysis of gfr slope as a surrogate endpoint for kidney failure.
\newblock {\em Nature Medicine}, pages 1--10.

\bibitem[Inker et~al., 2019]{inker2019gfr}
Inker, L.~A., Heerspink, H.~J., Tighiouart, H., Levey, A.~S., Coresh, J., Gansevoort, R.~T., Simon, A.~L., Ying, J., Beck, G.~J., Wanner, C., et~al. (2019).
\newblock Gfr slope as a surrogate end point for kidney disease progression in clinical trials: a meta-analysis of treatment effects of randomized controlled trials.
\newblock {\em Journal of the American Society of Nephrology: JASN}, 30(9):1735.

\bibitem[{International Council for Harmonisation of Technical Requirements for Pharmaceuticals for Human Use}, 2021]{ich-e9-r1}
{International Council for Harmonisation of Technical Requirements for Pharmaceuticals for Human Use} (2021).
\newblock Statistical principles for clinical trials: Addendum: Estimands and sensitivity analysis in clinical trials.
\newblock Guideline.
\newblock E9(R1).

\bibitem[Jha et~al., 2013]{jha2013chronic}
Jha, V., Garcia-Garcia, G., Iseki, K., Li, Z., Naicker, S., Plattner, B., Saran, R., Wang, A. Y.-M., and Yang, C.-W. (2013).
\newblock Chronic kidney disease: global dimension and perspectives.
\newblock {\em The Lancet}, 382(9888):260--272.

\bibitem[Khan et~al., 2022]{khan2022potential}
Khan, M.~S., Bakris, G.~L., Shahid, I., Weir, M.~R., and Butler, J. (2022).
\newblock Potential role and limitations of estimated glomerular filtration rate slope assessment in cardiovascular trials: a review.
\newblock {\em JAMA cardiology}, 7(5):549--555.

\bibitem[Levey et~al., 2020]{levey2020change}
Levey, A.~S., Gansevoort, R.~T., Coresh, J., Inker, L.~A., Heerspink, H.~L., Grams, M.~E., Greene, T., Tighiouart, H., Matsushita, K., Ballew, S.~H., et~al. (2020).
\newblock Change in albuminuria and gfr as end points for clinical trials in early stages of ckd: a scientific workshop sponsored by the national kidney foundation in collaboration with the us food and drug administration and european medicines agency.
\newblock {\em American journal of kidney diseases}, 75(1):84--104.

\bibitem[Perkovic et~al., 2019]{perkovic2019canagliflozin}
Perkovic, V., Jardine, M.~J., Neal, B., Bompoint, S., Heerspink, H.~J., Charytan, D.~M., Edwards, R., Agarwal, R., Bakris, G., Bull, S., et~al. (2019).
\newblock Canagliflozin and renal outcomes in type 2 diabetes and nephropathy.
\newblock {\em New England journal of medicine}, 380(24):2295--2306.

\bibitem[Perkovic et~al., 2024]{perkovic2024effects}
Perkovic, V., Tuttle, K.~R., Rossing, P., Mahaffey, K.~W., Mann, J.~F., Bakris, G., Baeres, F.~M., Idorn, T., Bosch-Traberg, H., Lausvig, N.~L., et~al. (2024).
\newblock Effects of semaglutide on chronic kidney disease in patients with type 2 diabetes.
\newblock {\em New England Journal of Medicine}, 391(2):109--121.

\bibitem[Stevens et~al., 2024]{stevens2024kdigo}
Stevens, P.~E., Ahmed, S.~B., Carrero, J.~J., Foster, B., Francis, A., Hall, R.~K., Herrington, W.~G., Hill, G., Inker, L.~A., Kazanc{\i}o{\u{g}}lu, R., et~al. (2024).
\newblock Kdigo 2024 clinical practice guideline for the evaluation and management of chronic kidney disease.
\newblock {\em Kidney international}, 105(4):S117--S314.

\bibitem[Thompson et~al., 2020]{thompson2020change}
Thompson, A., Smith, K., and Lawrence, J. (2020).
\newblock Change in estimated gfr and albuminuria as end points in clinical trials: a viewpoint from the fda.
\newblock {\em American Journal of Kidney Diseases}, 75(1):4--5.

\bibitem[Tuttle et~al., 2018]{tuttle2018dulaglutide}
Tuttle, K.~R., Lakshmanan, M.~C., Rayner, B., Busch, R.~S., Zimmermann, A.~G., Woodward, D.~B., and Botros, F.~T. (2018).
\newblock Dulaglutide versus insulin glargine in patients with type 2 diabetes and moderate-to-severe chronic kidney disease (award-7): a multicentre, open-label, randomised trial.
\newblock {\em The lancet Diabetes \& endocrinology}, 6(8):605--617.

\bibitem[Vonesh et~al., 2019]{vonesh2019mixed}
Vonesh, E., Tighiouart, H., Ying, J., Heerspink, H.~L., Lewis, J., Staplin, N., Inker, L., and Greene, T. (2019).
\newblock Mixed-effects models for slope-based endpoints in clinical trials of chronic kidney disease.
\newblock {\em Statistics in medicine}, 38(22):4218--4239.

\bibitem[Vonesh and Greene, 2022]{vonesh2022biased}
Vonesh, E.~F. and Greene, T. (2022).
\newblock Biased estimation with shared parameter models in the presence of competing dropout mechanisms.
\newblock {\em Biometrics}, 78(1):399--406.

\bibitem[Webster et~al., 2017]{webster2017chronic}
Webster, A.~C., Nagler, E.~V., Morton, R.~L., and Masson, P. (2017).
\newblock Chronic kidney disease.
\newblock {\em The lancet}, 389(10075):1238--1252.

\end{thebibliography}

\end{document}